\documentclass{article}

\usepackage[final]{airov26}

\usepackage{graphicx}
\usepackage{subcaption}

\usepackage[utf8]{inputenc} 
\usepackage[T1]{fontenc}    
\usepackage{booktabs}       
\usepackage{amsmath, amsfonts}       
\usepackage{nicefrac}       
\usepackage{microtype}      
\usepackage{xcolor}         
\usepackage{threeparttable}
\usepackage{float}

\title{Linearized Bregman Iterations for Sparse Spiking Neural Networks}

\newcommand{\figref}[1]{Fig.~\ref{#1}}

\newcommand{\tabref}[1]{Tab.~\ref{#1}}
\author{%
	Daniel Windhager \\
	Silicon Austria Labs \\
	Linz, Austria \\
	\texttt{daniel.windhager@silicon-austria.com} \\
	\And
	Bernhard A. Moser\thanks{ double affiliation with Institute of Signal Processing, Johannes Kepler University Linz} \\
	Software Competence Center Hagenberg\\
	Hagenberg, Austria\\   
	\texttt{bernhard.moser@scch.at} \\
	\And
	Michael Lunglmayr\\
	Institute of Signal Processing\\
	Johannes Kepler University \\
	Linz, Austria\\
	\texttt{michael.lunglmayr@jku.at} \\
}

\begin{document}

\maketitle

\begin{abstract}
Spiking Neural Networks (SNNs) offer an energy efficient alternative to conventional Artificial Neural Networks (ANNs) but typically still require a large number of parameters. This work introduces \emph{Linearized Bregman Iterations} (LBI) as an optimizer for training SNNs, enforcing sparsity through iterative minimization of the Bregman distance and proximal soft thresholding updates. To improve convergence and generalization, we employ the \emph{AdaBreg} optimizer, a momentum and bias corrected Bregman variant of Adam. Experiments on three established neuromorphic benchmarks, i.e. the Spiking Heidelberg Digits (SHD), the Spiking Speech Commands (SSC), and the Permuted Sequential MNIST (PSMNIST) datasets, show that LBI based optimization reduces the number of active parameters by about $50\%$ while maintaining accuracy comparable to models trained with the Adam optimizer, demonstrating the potential of convex sparsity inducing methods for efficient neuromorphic learning.
\end{abstract}

\section{Introduction}
Sparsity in neural networks is an important and ongoing research field \cite{hoefler2021sparsity}. In most neural network architectures sparsity refers to the weight matrices of the model being sparsely populated. This has the effect of sparsifying connections of the network (i.e. a connection with weight $0$ is disconnected). For classical GPU implementations, however, having sparse weights can sometimes actually lead to increased computational overhead due to memory organization problems and unbalanced workloads \cite{zaharia2020sparse}. In contrast, for edge AI implementations where network structures are directly mapped to hardware, e.g. as in \cite{mineuron}, the benefits are considerably higher as the sparsity of weights can be utilized and exploited more easily. 

Training machine learning to produce sparse weight matrices is in essence a sparse optimization problem. In practice, sparsity is often achieved through heuristic approaches such as pruning \cite{hoefler2021sparsity}, even though there exist mathematically founded algorithms that can provably converge to sparse optimal solutions. A number of these algorithms is based on so-called linearized Bregman iterations, which have also been proposed for sparse estimation \cite{osher2005iterative, yin2008bregman} using concepts similar to Douglas–Rachford splitting \cite{douglasRachford}. Their efficiency and stability for sparse estimation have been demonstrated repeatedly \cite{OBLI, EUROCAST17}, and more recent work has shown their suitability for training sparse deep neural networks, in many cases outperforming heuristic solutions \cite{Bungert}.

In this work, we investigate how linearized-Bregman-based sparse learning performs for spiking neural networks. Specifically, we evaluate both feedforward and recurrent SNN architectures on established neuromorphic benchmarks and analyze how the regularization parameter $\lambda$ influences sparsity and model accuracy.

\section{Linearized Bregman Iterations}
\label{s:LBI}
Sparse optimization problems often include a regularization term \( J(\boldsymbol{\theta}) \) based on the \(\ell_1\)-norm to promote sparsity. However, the non-smoothness of the \(\ell_1\)-norm causes standard gradient-based optimization to fail in regions where the gradient is undefined. A key property that enables a well-behaved optimization framework in such cases is the \emph{convexity} of the regularization function \(J\). For convex but possibly non-smooth functions, the concept of a gradient is replaced by a \emph{sub-differential}, which generalizes differentiation to non-smooth settings. Sub-differentials, however, can not directly be used in steepest-decent like algorithms as they require a defined gradient at each step.

To handle such cases, Bregman iterations iteratively minimize the \emph{Bregman distance} \cite{bregman1967relaxation}
\[
D_J(x, y) = J(x) - J(y) - \langle \nabla J(y), x - y \rangle,
\]
rather than minimizing the composite cost function directly. In this formulation, \(\nabla J(y)\) represents an element of the \emph{sub-differential} of \(J\) at point \(y\), denoted \(\partial J(y)\). For convex \(J\), the sub-differential \(\partial J(y)\) is a non-empty, convex set that captures all possible slopes of local supporting hyperplanes to \(J\) at \(y\). For example, when \(J(x) = |x|\), the sub-differential at \(x = 0\) is the interval \([-1, 1]\). This interpretation allows the Bregman distance to generalize classical gradient-based methods to convex but non-differentiable regularizers, enabling optimization for the \(\ell_1\)-norm and similar sparsity-promoting terms.

Because the resulting subproblems rarely admit closed-form solutions for sparse regularizers \cite{yin2008bregman}, the \emph{Linearized Bregman Iteration} (LBI) method provides an efficient linearized approximation. LBI introduces auxiliary ``shadow'' variables \({\bf v}\) corresponding to the model parameters \(\boldsymbol{\theta}\). At iteration \(t\), the updates can be expressed as
\[
{\bf v}^{(t+1)} = {\bf v}^{(t)} + \mu \nabla L(\boldsymbol{\theta}^{(t)}, B),
\qquad
\boldsymbol{\theta}^{(t+1)} = \text{prox}_{J}({\bf v}^{(t+1)}),
\]
where \(\mu\) denotes the step size and \(\text{prox}_J\) represents the proximal operator associated with the convex regularization term \(J\).

For the commonly used sparse regularizer \( J(\boldsymbol{\theta}) = \lambda \|\boldsymbol{\theta}\|_1 \), the proximal operator corresponds to the elementwise \emph{soft-thresholding} function,
\begin{align}
\text{prox}_{\lambda\|\boldsymbol{\theta}\|_1}(\boldsymbol{\theta}) &= \left[ \text{prox}_{\lambda\|\boldsymbol{\theta}\|_1}(\theta_i)\right]_i\\
\text{prox}_{\lambda\|\boldsymbol{\theta}\|_1}(\theta_i) &= \text{sign}(\theta_i)\,\max(0, |\theta_i| - \lambda),
\end{align}
which suppresses small weight values and drives many parameters exactly to zero, thereby yielding sparse network representations.

This mechanism makes LBI particularly well suited for training models such as Spiking Neural Networks, where convex sparse regularization aligns well with the need for energy-efficient, low-parameter inference on neuromorphic hardware.

\subsection{AdaBreg Optimization}
\label{ss:AdaBreg}
While classical Linearized Bregman Iterations provide an effective framework for promoting sparsity, their convergence can be slow when applied to large-scale neural network training. To improve adaptation and generalization, we adopt the AdaBreg optimizer introduced by \cite{Bungert}, which extends the Bregman iteration concept by incorporating adaptive moment estimation in analogy to the Adam optimizer \cite{adam}.

AdaBreg inherits the shadow-variable formulation of the linearized Bregman framework while maintaining separate exponential moving averages of the first and second moments of the gradient. Let \(\mathbf{m}_t\) and \(\mathbf{s}_t\) denote the biased estimates of the mean and variance of the stochastic gradient \(\nabla L(\boldsymbol{\theta}_t, B)\) at iteration \(t\). For \(t=0\), both \(\mathbf{m}_t\) and \(\mathbf{s}_t\) may be set to \(\boldsymbol{0}\) and \(0\) respectively. The update rules can then be summarized as
\[
\begin{aligned}
\mathbf{m}_{t+1} &= \beta_1 \mathbf{m}_t + (1 - \beta_1) \nabla L(\boldsymbol{\theta}_t, B), \\
\mathbf{s}_{t+1} &= \beta_2 \mathbf{s}_t + (1 - \beta_2) \big(\nabla L(\boldsymbol{\theta}_t, B)\big)^2, \\
\mathbf{v}_{t+1} &= \mathbf{v}_t + \mu \frac{\widehat{\mathbf{m}}_{t+1}}{\sqrt{\widehat{\mathbf{s}}_{t+1}} + \epsilon}, \\
\boldsymbol{\theta}_{t+1} &= \text{prox}_{\lambda\|\boldsymbol{\theta}\|_1}(\mathbf{v}_{t+1}),
\end{aligned}
\]
where \(\mu\) denotes the learning rate, \(\widehat{\mathbf{m}}_{t+1}\) and \(\widehat{\mathbf{s}}_{t+1}\) represent bias-corrected moment estimates, and \(\epsilon\) is a small numerical constant for stability.


In practice, the optimizer can be seamlessly integrated into existing deep learning frameworks such as PyTorch by substituting the Adam optimizer with its AdaBreg counterpart, requiring no structural changes to the network implementation.

\section{Results}
\subsection{Setup and configurations}
All of the results presented in this work were obtained by training the networks using the AdaBreg algorithm introduced by \cite{Bungert}, which is a Bregman version of the Adam algorithm \cite{adam} that includes momentum and a bias correction term. AdaBreg combines the adaptive moment estimation of Adam with the sparsity inducing properties of Linearized Bregman Iterations, enabling direct integration of convex regularization into the optimization process. The reason for choosing AdaBreg over the standard linearized Bregman iteration based algorithm with momentum (LinBreg) is the better performance and generalization capability as reported by the original authors. Unless otherwise stated, the same sets of hyperparameters and initialization conditions were used across all experiments for comparability.

For the datasets, three of the most common datasets among the neuromorphic community were chosen, namely Spiking Heidelberg Digits (SHD), Spiking Speech Commands (SSC) and the Permuted Sequential MNIST (PSMNIST) dataset, along with the neuron network models presented by \cite{delrec}, who at the time of writing hold the record for the top performing SNN on the SSC dataset. These datasets jointly cover a broad range of temporal complexities, from short event based auditory signals (SHD) to long sequential patterns (PSMNIST), providing a balanced testbed for evaluating sparsity effects across different task domains.

\begin{table}[ht]
    \centering
    \renewcommand{\arraystretch}{1.3}
    \begin{threeparttable}
        \caption{Number of neurons and layers of the networks used for evaluations of Linearized Bregman iterations on the SHD, SSC and PSMNIST datasets.}
        \begin{tabular}{lccccc}
         \hline
         Dataset & Inputs & Hidden Layer 1 & Hidden Layer 2 & Hidden Layer 3 & Outputs \\
         \hline
         SHD     & $140^\S$  & $256$          & $256^\dag$     & -              & $20$ \\
         SSC     & $140^\S$  & $256^\dag$     & $256^\dag$     & $256^\dag$     & $35\ddag$ \\
         PSMNIST & $1$    & $64^\dag$      & $212^\dag$     & $212^\dag$     & $10\ddag$ \\
         \hline
    \end{tabular}
        
    \begin{tablenotes}
        \item[$\S$] Inputs are reduced from the original 700 inputs, by binning with a factor of 5. 
        \item[$\dag$] Recurrent layer with axonal delays in the recurrent path.
        \item[$\ddag$] Outputs from last linear layer are used directly, without LIF activation.
    \end{tablenotes}
        
    \label{tab:networks}
    \end{threeparttable}
\end{table}

The networks chosen for the datasets consist of either three or four layers with different feature sizes and configurations. All networks use the Leaky Integrate and Fire (LIF) neuron model. For the SHD dataset, the first and last layers of the network are simple feedforward SNN layers without any delay, while the middle layer is a recurrent layer with learned axonal delays in the recurrent path. The networks for the SSC and PSMNIST datasets each consist of three recurrent SNN layers with learned axonal delays in their recurrent paths, followed by a final linear layer without LIF activation. A simplified overview of the networks can be seen in \tabref{tab:networks}, while further architectural details, including delay learning mechanisms can be found in \cite{delrec}.

\subsection{Performance with learning rate schedulers}
\label{sec:perf_lr}
\begin{figure}[htbp]
    \centering
    \begin{subfigure}[t]{0.49\textwidth}
        \centering
        \includegraphics[width=\textwidth]{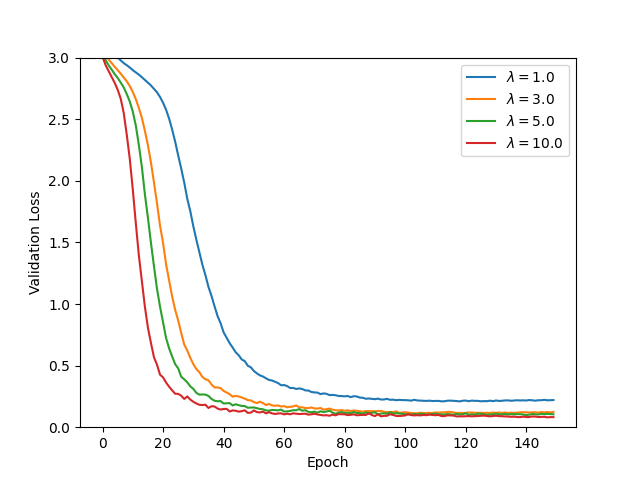}
        \caption{Loss curve for SHD dataset, when trained with the OneCycleLR scheduler from PyTorch for 150 epochs, with an initial learning rate of $5\cdot{}10^{-3}$.}
        \label{fig:1a}
    \end{subfigure}
    \hfill 
    \begin{subfigure}[t]{0.49\textwidth}
        \centering
        \includegraphics[width=\textwidth]{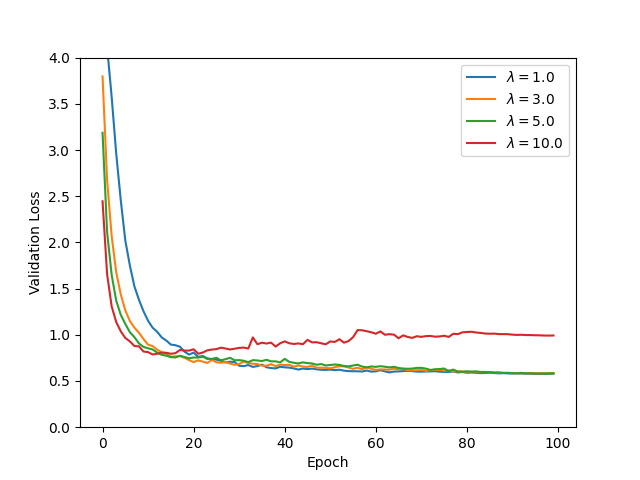}
        \caption{Loss curve for SSC dataset, when trained with the OneCycleLR scheduler from PyTorch for 100 epochs, with an initial learning rate of $1\cdot{}10^{-3}$. Onset of divergent behaviour can be seen for $\lambda=10$.}
        \label{fig:1b}
    \end{subfigure}
    \par\bigskip
    \begin{subfigure}[t]{0.49\textwidth}
        \centering
        \includegraphics[width=\textwidth]{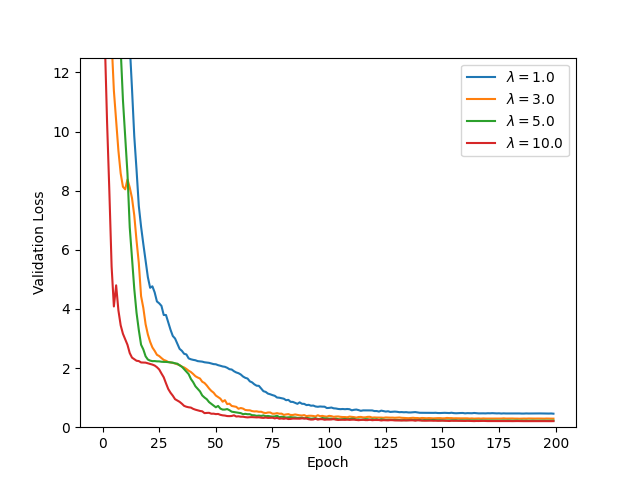}
        \caption{Loss curve for PSMNIST dataset, when trained with the OneCycleLR scheduler from PyTorch for 200 epochs, with an initial learning rate of $1\cdot{}10^{-3}$.}
        \label{fig:1c}
    \end{subfigure}
    
    \caption{Loss curves for the training on SHD, SSC and PSMNIST dataset with different values for $\lambda$. All curves were averaged over three separate training runs with different seeds.}
    \label{fig:loss}
\end{figure}

The loss on the validation sets can be seen in \figref{fig:loss}, with the different colored curves indicating various values for the parameter $\lambda$, which controls the sparsity of the solution. These results indicate that larger $\lambda$ values also introduce a beneficial regularization effect, leading to faster initial convergence and smoother loss trajectories at the early stages of training. This behavior is consistent across datasets and highlights the dual role of $\lambda$ as both a sparsity and regularization parameter. 

LBI seems to be more sensitive to higher learning rates, especially when used in conjunction with learning rate schedulers, which might be caused by the inherent stagnation phases that occur during training with Linearized Bregman iterations. In fact, choosing a high enough learning rate, paired with large values of the sparsity controlling parameter $\lambda$, causes the training to diverge after a few epochs. The onset of this divergent behaviour can be seen in \figref{fig:1b} for $\lambda=10$, which is due to $\lambda$ being slightly too high for the chosen learning rate.

\begin{figure}[t]
    \centering
    \begin{subfigure}[t]{0.49\textwidth}
        \centering
        \includegraphics[width=\textwidth]{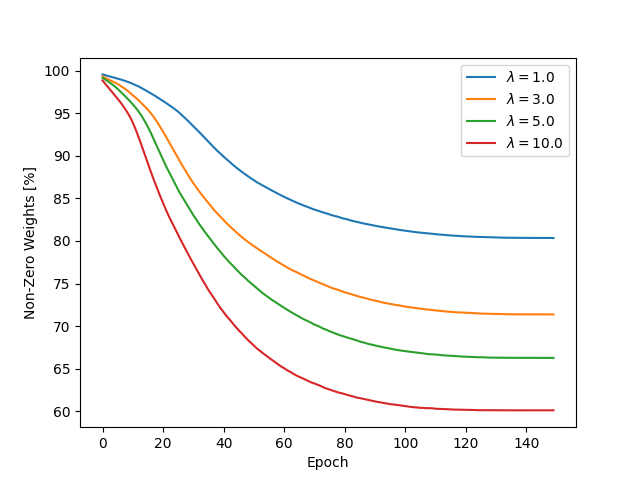}
        \caption{Progression of number of non-zero weights for the neural network trained on the SHD dataset.}
        \label{fig:2a}
    \end{subfigure}
    \hfill 
    \begin{subfigure}[t]{0.49\textwidth}
        \centering
        \includegraphics[width=\textwidth]{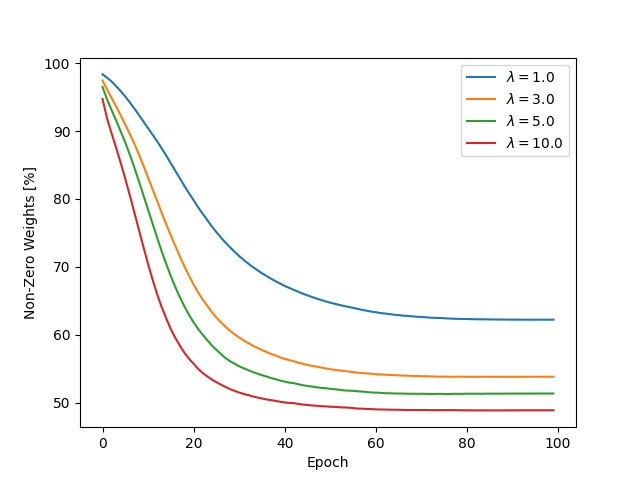}
        \caption{Progression of number of non-zero weights for the neural network trained on the SSC dataset.}
        \label{fig:2b}
    \end{subfigure}
    \par\bigskip
    \begin{subfigure}[t]{0.49\textwidth}
        \centering
        \includegraphics[width=\textwidth]{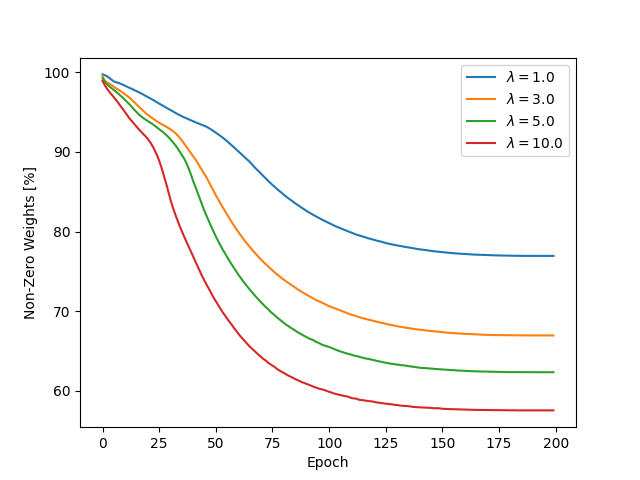}
        \caption{Progression of number of non-zero weights for the neural network trained on the PSMNIST dataset.}
        \label{fig:2c}
    \end{subfigure}
    
    \caption{Number of non-zero values in networks for SHD, SSC and PSMNIST datasets during the training process plotted as a function of current epoch for different $\lambda$ values. Results represent the mean across three independent training runs.}
    \label{fig:sparsity}
\end{figure}

The goal of Linearized Bregman training is to produce highly sparse weight matrices, i.e. containing a large fraction of zero entries. As can be seen in \figref{fig:sparsity}, the number of non-zero parameters across the entire network does indeed decrease monotonically as training progresses. Much like the loss curves, the sparsity level increases rapidly during early training epochs before reaching a plateau, a behavior characteristic of Bregman iterations which preferentially eliminate less important features in the initial optimization phases.

\begin{figure}
    \centering
    \includegraphics[width=0.5\linewidth]{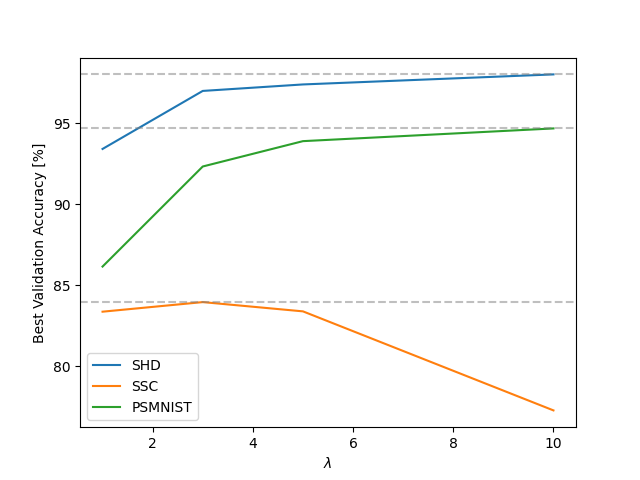}
    \caption{Peak validation accuracy across SHD, SSC, and PSMNIST datasets as a function of regularization parameter $\lambda$, averaged over multiple training runs. The optimal $\lambda$ is slightly higher for SHD and SSC, while a lower value of $\lambda$ achieves the best results for PSMNIST.}
    \label{fig:val_acc_over_lambda}
\end{figure}

Since the parameter $\lambda$ clearly influences both the best achieved accuracy and the achieved sparsity of the network, a natural question is the optimal selection of $\lambda$. \figref{fig:val_acc_over_lambda} shows the best validation accuracy achieved across multiple training runs for all three datasets, plotted as a function of the chosen $\lambda$ value. These results indicate that the choice of $\lambda$ must be made depending on the learning rate and the chosen dataset. For SHD and SSC a slightly higher value of $\lambda$ is beneficial, while it results in worse performance for the PSMNIST dataset.

\subsection{Performance without learning rate schedulers}
\begin{figure}[htbp]
    \centering
    \begin{subfigure}[b]{0.49\textwidth}
        \centering
        \includegraphics[width=\textwidth]{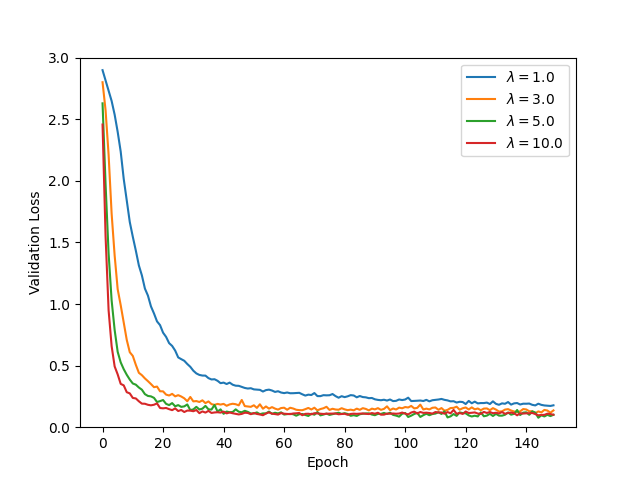}
        \caption{Loss curve for SHD dataset, when trained without a learning rate scheduler for 150 epochs, with an initial learning rate of $2\cdot{}10^{-4}$.}
        \label{fig:3a}
    \end{subfigure}
    \hfill 
    \begin{subfigure}[b]{0.49\textwidth}
        \centering
        \includegraphics[width=\textwidth]{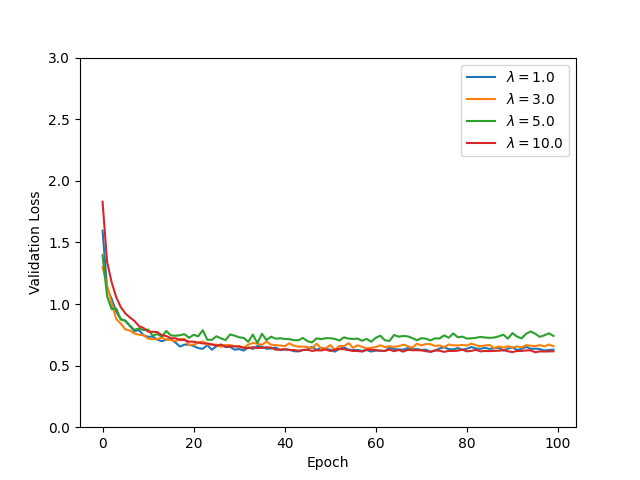}
        \caption{Loss curve for SSC dataset, when trained without a learning rate scheduler for 100 epochs, with an initial learning rate of $5\cdot{}10^{-4}$.}
        \label{fig:3b}
    \end{subfigure}
    \par\bigskip
    \begin{subfigure}[b]{0.49\textwidth}
        \centering
        \includegraphics[width=\textwidth]{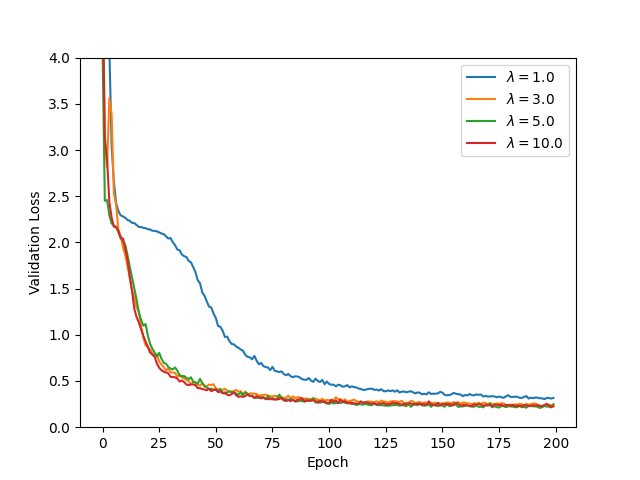}
        \caption{Loss curve for PSMNIST dataset, when trained without a learning rate scheduler for 200 epochs, with an initial learning rate of $1\cdot{}10^{-4}$.}
        \label{fig:3c}
    \end{subfigure}
    
    \caption{Loss curves for the training on SHD, SSC and PSMNIST dataset with different values for $\lambda$. All curves represent means over three independent training runs with different random seeds.}
    \label{fig:loss_no_lr}
\end{figure}

Since learning rate schedulers impact the training process and can even cause training divergence when coupled with a high enough learning rate, the previous experiments were repeated without any learning rate scheduling during the training. The resulting loss curves can be seen in \figref{fig:loss_no_lr}. 

From the subfigures in \figref{fig:loss_no_lr} it is apparent that the achieved accuracy is largely unaffected by the presence or absence of learning rate schedulers, further substantiating the hypothesis that the initial divergent behaviour was due to a too high learning rate. Furthermore, the number of non-zero weights in the network, i.e., the sparsity level, also remained largely unaffected by the presence or absence of schedulers. These sparsity curves are therefore omitted here for conciseness, as they closely mirror those shown in \figref{fig:sparsity}.

\begin{figure}[h]
    \centering
    \includegraphics[width=0.5\linewidth]{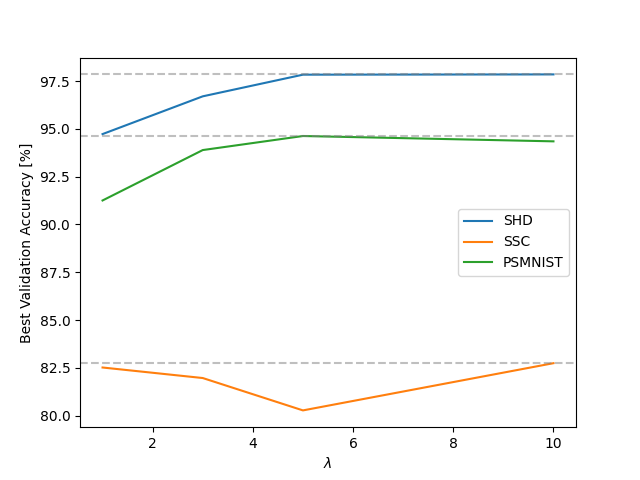}
    \caption{Peak validation accuracy without learning rate scheduling across SHD, SSC, and PSMNIST datasets versus regularization parameter $\lambda$, averaged over multiple training runs.}
    \label{fig:val_acc_over_lambda_no_lr}
\end{figure}

The achieved best validation accuracy across all datasets without learning rate scheduling, plotted as a function of the chosen $\lambda$ value, can be seen in \figref{fig:val_acc_over_lambda_no_lr}. These results demonstrate that without learning rate scheduling, a higher value of $\lambda$ may be chosen sometimes (see results for PSMNIST in \figref{fig:val_acc_over_lambda}), thus potentially increasing regularization and performance on unseen data, although the effect seems comparatively small when comparing the performance on the test sets.

\section{Performance compared to training with Adam}
The baseline models from \cite{delrec} were trained using the well-known Adam optimizer \cite{adam}. A direct comparison of test set performance between their results and ours is provided in \tabref{tab:performance}. This comparison reveals that Linearized Bregman iterations (AdaBreg) achieve accuracies within 0.5--1.5\% of the Adam baseline across all three datasets, despite limited hyperparameter tuning. When combined with the observed $\approx 50\%$ reduction in active parameters (cf.\ \figref{fig:sparsity}), this performance gap appears acceptable for sparsity-constrained neuromorphic applications.

\renewcommand{\arraystretch}{1.3}
\begin{table}[h]
    \centering
    \caption{Test set accuracy comparison. Baseline results from \cite{delrec} (Adam optimizer) versus AdaBreg results with and without learning rate scheduling across SHD, SSC, and PSMNIST datasets.}
    \vspace{2mm}
    \begin{tabular}{lccc}
        \hline
         Dataset                 & SHD   & SSC   & PSMNIST \\\hline
         \cite{delrec}           & 93.39\% & 82.58\% & 96.21\%   \\
         Ours                    & 92.98\% & 81.86\% & 95.59\%   \\
         Ours (no LR scheduling) & 92.28\% & 80.67\% & 95.11\%   \\
         \hline
    \end{tabular}
    \label{tab:performance}
\end{table}

\section{Conclusion}
This work demonstrates the practical viability of Linearized Bregman Iterations (LBI) as an optimizer for Spiking Neural Networks (SNNs). Across three established neuromorphic benchmarks (SHD, SSC, PSMNIST), LBI-based training (AdaBreg) achieves competitive accuracy within 0.5--1.5\% of Adam baselines while reducing the number of active parameters by approximately 50\%.

Further performance gains appear achievable through systematic hyperparameter optimization. Notably, LBI integrates seamlessly into existing PyTorch workflows—requiring only a one-line optimizer replacement (Adam $\rightarrow$ AdaBreg)—dramatically lowering the adoption barrier for sparsity-aware SNN training.

These findings highlight a critical hardware-software co-design opportunity: while sparse SNN training is now readily accessible, neuromorphic hardware must evolve to fully exploit this parameter efficiency for energy-constrained edge deployments.

\section*{Acknowledgements}
This work was supported by (1) the ’University SAL Labs’ initiative of Silicon Austria Labs (SAL) and its Austrian partner universities for applied fundamental research for electronic based systems, and (2) the COMET Programme via SCCH funded by the Austrianministries BMIMI, BMWET, and the State of UpperAustria, (3) the COMET-K2 ``Center for Symbiotic Mechatronics'' of the Linz Center of Mechatronics (LCM), funded by the Austrian federal government and the federal state of Upper Austria.

The research reported in this paper has also been partly funded by the European Union’s Horizon 2020 research and innovation program within the framework of Chips Joint Undertaking (Grant No. 101112268). This work has been supported by Silicon Austria Labs (SAL) owned by the Republic of Austria, the Styrian Business Promotion Agency (SFG), the federal state of Carinthia, the Upper Austrian Research (UAR), and the Austrian Association for the Electric and Electronics Industry (FEEI).

\bibliographystyle{abbrvnat}
\bibliography{bibliography}

\end{document}